# Graphitic nanofibres from electrospun solutions of PAN in dimethylsulphoxide

Zeynep Kurban[*a,b], Arthur Lovell[a,b], Derek Jenkins[c], Steve Bennington[a,b], Ian Loader[c], Alex Schober[a], Neal Skipper[b]

[a]ISIS Facility, STFC Rutherford Appleton Laboratory, Didcot, OX11 0QX, UK

[b]Department of Physics and Astronomy, UCL, Gower Street, London, WC1E 6BT, UK

[c]Micro and Nanotechnology Centre, STFC Rutherford Appleton Laboratory, Didcot, OX11 0QX, UK

## Abstract

Homogenous graphitic nanofibres (GNFs) have been synthesised by heat treatment of electrospun polyacrylonitrile in dimethylsulphoxide, offering a new solution route of low toxicity to manufacture sub-60 nm diameter GNFs. Fibre beading resulting from the spinning of low-concentration polymer solutions can be reduced with the addition of surfactant or sodium chloride. Characterisation techniques including X-ray diffraction, scanning- and transmission electron microscopy have been used to quantify the effect of the graphitisation process, by heat treatment up to 3000°C, on the weight, diameter and structural morphology of the nanofibres. The GNFs have an entangled micro-fibril structure with graphitic ordering of up to 40 graphene layers after treatment at 3000°C. There is little difference in the degree of graphitisation of GNFs prepared with a 250°C oxidation step compared with those prepared without, but oxidised GNFs retain more of their original mass after heating under argon flow.

---

[*] Corresponding author. Email: zeynep.kurban@stfc.ac.uk Fax: (+44) 1235 445720

## 1. Introduction

Carbon fibres have been historically used in applications from sports equipment to the aerospace industry. Their high tensile strength [1, 2], large length-to-diameter ratio[3], high specific surface area[4] and high thermal and electrical conductivity [5, 6] have made them increasingly popular materials to investigate for new applications, utilising improved production methods capable of producing large quantities of nanoscale fibres. Among other potential applications for carbon nanofibres (CNFs) are templates for nanotubes [7], filters [8], supercapacitors [9], batteries [10], and bottom-up assembly in nanoelectronics [9] and photonics [11].

In this study we report and characterise CNFs synthesised by heat treatment from polymer fibres manufactured using an electrospinning process. The resulting fibres are graphitised up to 3000°C to make high specific surface area [12] (SSA) graphitic nanofibres, which by modification of pore diameters and interlayer spacing are of interest for high-density storage of hydrogen.

The electrospinning method [13, 14] has been known for over a century [15] but has only recently begun to displace traditional CNF synthesis methods such as vapour growth, plasma-enhanced chemical vapour deposition methods, wet spinning and stretching methods [16]. We have used electrospinning [17-19] because it can produce homogenous fibres with diameters of 10-400 nm and kilometre lengths, in a scalable and efficient way [14].

In the archetypal process, a solution of a polymeric precursor in a suitable organic solvent is extruded electrostatically from a charged metal nozzle onto an earthed collector, usually a metal plate or drum. The distance between nozzle and collector is sufficient that most of the solvent evaporates in the precipitation process, while polymer entanglement allows a single very long fibre to be collected. The fibre length is a function of the stability of the spinning process and can be many hundreds of metres. The fibre diameter, conversely, is dependent on a large number of interdependent spinning parameters including: the average polymer chain length, the concentration of polymer in solvent, the distance between nozzle and collector, the electric field strength and shape, the flow rate of the solution through the nozzle, and the electromechanical properties of the solution [14]. Of the various precursors for making carbon fibres, polyacrylonitrile (PAN) has been the most commonly used, owing to its high carbon content. PAN-based CNFs with a minimum average fibre diameter of ~250nm are generally reported, with Li *et al.* recently reporting average diameters of 150 nm [5]. In the present work, it is shown that solutions of PAN in dimethylsulphoxide (DMSO) can produce polymer fibres of ~100 nm diameter, which following heating produce GNFs with ~50 nm diameter, comparable with the finest fibres reported using other solvents such as dimethylformamide (DMF) [1, 20-22]. DMSO has the advantage of relative environmental and physiological benignity over the more commonly used teratogenic DMF. As we are interested in minimising fibre diameter to make high-SSA GNFs for hydrogen adsorption studies, we also report here the quantitative tuning effect on fibre synthesis and diameter of the addition of surfactant, higher-conductivity polymer, and salt to the PAN/DMSO solution.

The synthesis of carbon fibres from polymer fibres generally involves three steps: i) stabilisation, ii) carbonisation and iii) graphitisation. The stabilisation process, in which the polymer is heated to 200-300°C in an oxygen-containing atmosphere, stabilises the oriented structure and prevents fibre fusion at higher temperatures during the treatment process of the fibres. This prevents chain scission and hence mass loss that usually occurs if fibres are heated in an inert atmosphere without stabilisation, and increases the fibres' tensile strength. Details of the mechanisms of this process can be found in references [23-25].

The carbonisation and the subsequent graphitisation steps are carried out from 500-3000°C, in an inert atmosphere or vacuum. After carbonisation at 500°C, the fibre diameter is reduced through the expulsion of non-carbon elements, leaving an amorphous carbon structure. Subsequently, graphitisation occurs from ~1500 to 3000°C, with the nanocrystalline graphite dimensions increasing with heat treatment temperature (HTT). The HTT, the polymer used and the process condition employed in making the fibres are the most influential factors affecting the development of the structure. We aim to develop a highly crystalline fibre structure, suitable for intercalation, enabling high-SSA GNFs optimised for hydrogen adsorption. To this end, we studied GNFs produced without a stabilisation step (non-oxidised) to compare diameters and degree of graphitisation with those made with stabilisation (oxidised).

Structural characterisation of the nanofibres, before and after heat treatment, was made using X-ray diffraction (XRD) and transmission electron microscopy (TEM). Field-emission scanning electron microscopy (FE-SEM) was used to study the morphology of fibres and to obtain micrographs for determining average fibre diameter.

## 2. Experimental

### 2.1 Preparation, electrospinning and imaging of PAN fibres

Polymer solutions were made by mixing PAN powder (150,000g/mol. wt.) in >99.5 % purity DMSO (both from Sigma Aldrich) in varying proportions by weight to obtain solutions of concentrations in the range 2-10 wt % PAN. The solutions took approximately a day to achieve a clear-coloured homogeneous state, but this time could be reduced by immersion in an ultrasonic bath heated to 40-50°C. Sodium chloride (NaCl) and the surfactant 4-styrenesulphonic acid, sodium salt hydrate (SASH) (Sigma-Aldrich), were additionally added to some solutions at concentrations of 1 mg/ml and 5 mg/ml respectively, to investigate the effect of these reagents on polymer morphology and diameter. Salt increases the conductivity of the solution without adverse affect on the viscosity, while the surfactant has the effect of reducing the solution surface tension and increasing the conductivity. As a trial investigation PAN and polyaniline (PAni) were mixed in DMSO. PAni does not dissolve fully in DMSO, but electrospinning of the more homogenous, low-PAni concentration samples was possible.

A Brookfield (Programmable –II+) viscometer and a Jenway 4510 conductivity meter were used to measure viscosity and conductivity of the as-prepared solutions at room temperature (about 21°C) prior to electrospinning.

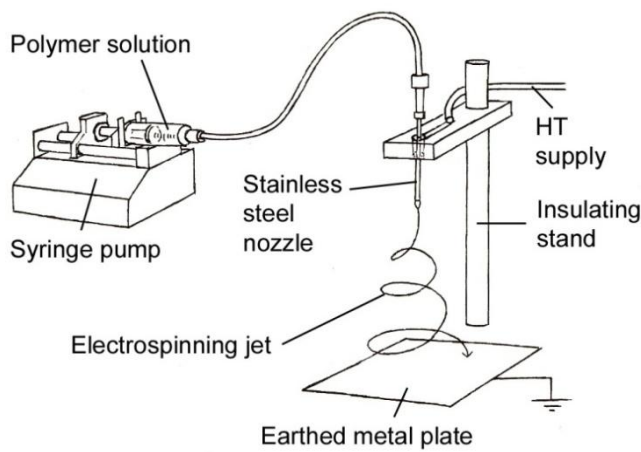

**Figure 1**: Electrospinning experimental apparatus, which is contained within a ventilated cabinet with access interlocked to the HT supply. Both syringe pump and HT supply are controllable from software. A closed-circuit camera (not shown) allows magnified observation of the nozzle tip for fine control of the spinning jet.

The experimental set-up of the electrospinning process, using the purpose built rig, is shown in Figure 1. The prepared solution was placed in a 10ml plastic syringe, connected to a stainless steel nozzle (0.5 mm inner diameter) fixed above an earthed aluminium collection plate. The feed rate of the solution was maintained with Harvard PHD 4400 programmable pump. The voltage supplied to the needle from a Glassman HT power supply was then adjusted until a stable spinning jet was established, as observed by a closed-circuit camera. This voltage varied between 5-10kV depending largely on the concentration and additives in the solution. To make the minimum diameter fibres the height of the nozzle above the Al collection plate was set to the maximum possible, 300 +/- 10 mm from nozzle tip to plate. After investigating the dependence of fibre diameters on flow rate, the lowest stable rate of 200ul/hr, which produced minimum diameter fibres, was used for the majority of the study. Fibres were collected on small Al plates (for microscopy studies) or graphite discs (for heat treatment) placed on the collection plate. The spinning jet was kept under control by observing the suspended droplet on the tip of the nozzle with the camera.

The fibres were imaged using a FE-SEM (Carl Zeiss XB1540) to determine the morphological appearance of the as-spun fibres. These SEM micrographs, taken from various regions on the SEM stub, were then used to determine average fibre diameters, using the image processing program *ImageJ*[1]. For each spinning condition, at least 100 and more typically around 300 readings for the fibre diameters were recorded. Statistical analysis of the data obtained was carried out by constructing a histogram, from which an arithmetic mean and a standard deviation were obtained.

## 2.2 Thermal treatment and characterisation of as-spun and treated PAN fibres

A Carbolite tube furnace was used to convert as-spun fibres into CNFs by heating at a ramp rate of 3°/min to 500°C, retained for one hour, in a glass tube under a dynamic vacuum. These carbonised fibres were then remeasured using SEM micrographs. The polymer solution used for the fibres that were graphitised at higher HTT was 6.8 % PAN in DMSO. This did not produce the finest fibres attained, but could produce more rapidly the larger fibre quantities required for XRD samples. The average fibre yield of this polymer solution when electrospun at 200 ul/hr was approximately 15 mg/hr, whereas for 4.2 % PAN in DMSO it was correspondingly less.

---

[1] Rasband WS. ImageJ. U. S. National Institutes of Health B, Maryland U, http://rsb.info.nih.gov/ij/, 1997-2008.

The graphitised PAN fibres were prepared with or without a stabilisation (oxidisation) step. While the non-oxidised fibres were simply carbonised to 500°C under vacuum for one hour as detailed above, the oxidised fibres were initially heat-treated in air at 250°C for six hours, following a temperature ramp of 3°/min, before the same carbonisation procedure. The resulting fibres were then heated in a Centorr Vacuum Systems Series 10 furnace to various heat treatment temperatures (HTT) from 1500 to 3000°C, retaining the 3°/min ramp rate and 1 hour plateau time.

The structure of the nanofibres was examined by X-ray diffraction (XRD), using a Philips X-Pert theta-theta X-ray diffractometer with Cu-Kα source, to determine the degree of graphitisation. A background obtained from an empty glass slide was subtracted from each pattern, and the intensities were normalised to the (100) peak. High magnification images of the fibres were also obtained using a TEM (JEOL JEM 2010) at an accelerating voltage of 200 kV, which also enabled a qualitative comparison with XRD of the extent of graphitisation of the fibres.

## 3. Results and Discussion

### 3.1 Polymer chemistry

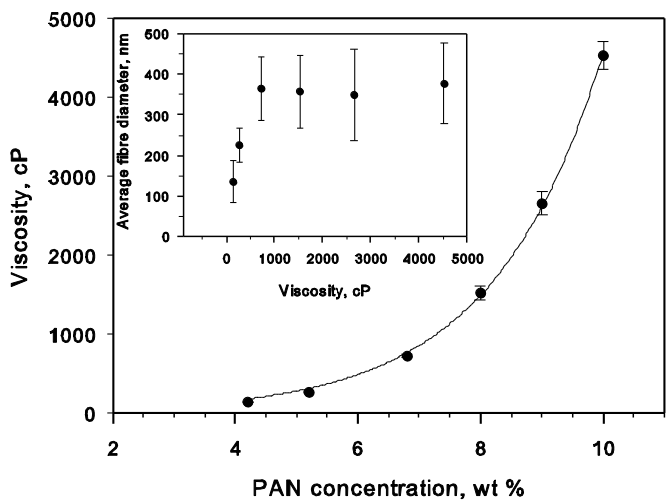

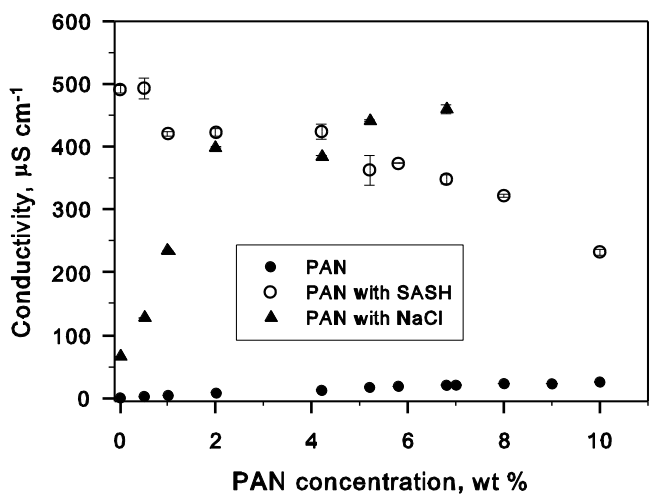

**Figure 2:** (a) Viscosity of PAN solutions in DMSO, the insert shows the fibre diameter as a function of viscosity; (b) Conductivity of PAN solutions in DMSO: (Closed circles) PAN in DMSO; (open circles) PAN in DMSO with 5 mg/ml 4-styrenesulphonic acid, sodium salt hydrate (SASH) surfactant; (triangles) PAN in DMSO with 1 mg/ml NaCl

The viscosity of pure PAN in DMSO solutions is shown in Figure 2(a). The curves of viscosity as a function of spindle speed were fitted to obtain a high-speed viscosity limit for each polymer concentration. This was plotted against PAN concentration and fitted to a positive exponential function. The increase in viscosity with increasing PAN concentration is clearly shown. The conductivity of pure PAN in DMSO with and without added SASH and NaCl is shown in Figure 2(b). Conductivity increases linearly with increased PAN concentration in the pure solution. Adding NaCl increases conductivity 20-fold up to 2 wt % PAN. The conductivity of SASH-containing solution decreases with PAN concentration. Figure 3 shows polymer fibres spun from 4.2 wt % PAN in DMSO, with and without SASH surfactant.

As for other polymer solutions, the diameter of spun fibres in a stable spinning state increases with solution flow rate and PAN concentration in PAN/DMSO solution. At low (≤5.2 wt %) PAN concentration, beaded fibres are formed as in Figure 3(a). The measured fibre diameters do not account for the beading, which appears to stretch out the fibres in between beads.

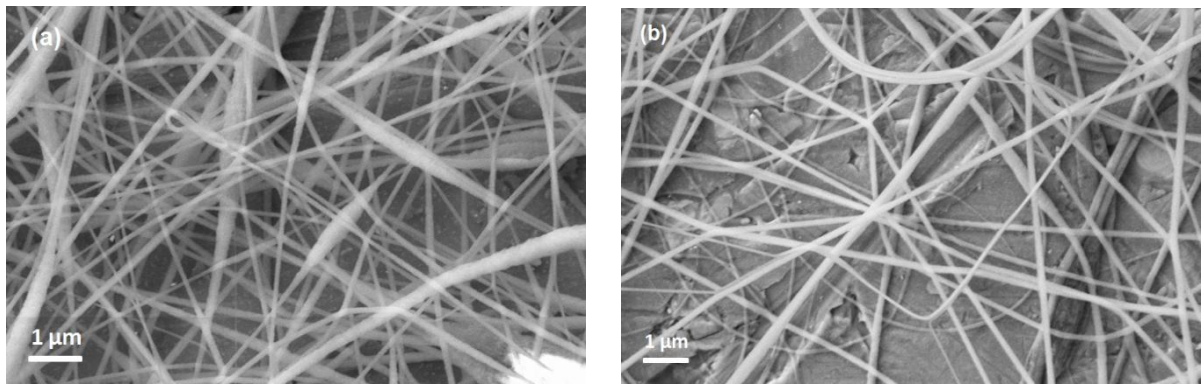

**Figure 3**: Fibres spun from 4.2 wt% PAN/DMSO solution; a) fibres without SASH; b) fibres with SASH

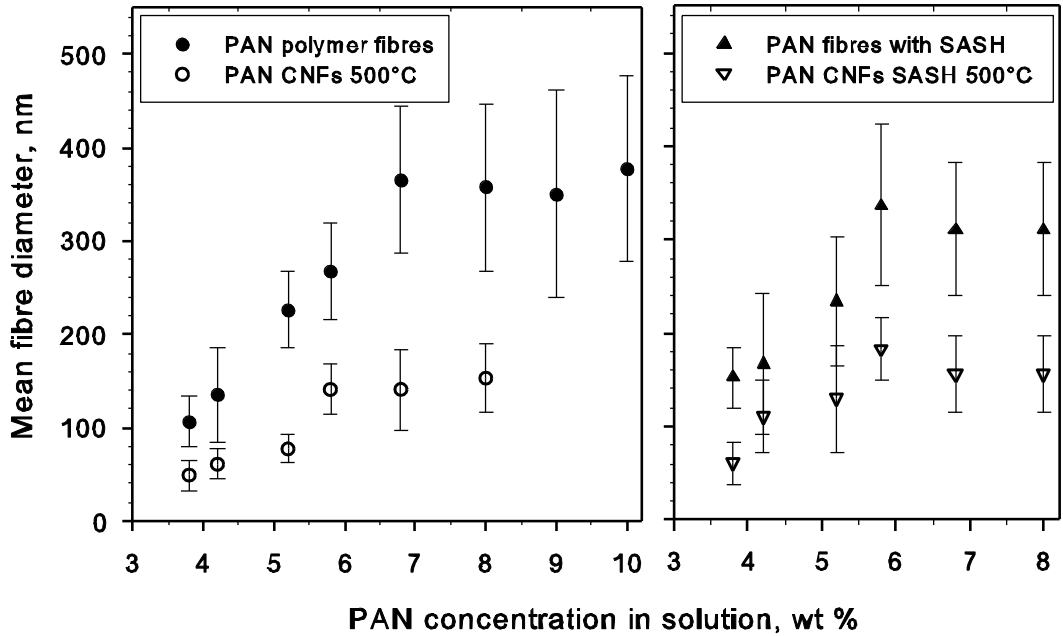

**Figure 4:** Polymer and 500°C pyrolysed (CNF) fibre diameters as a function of concentration of PAN in DMSO solution, without (left) and with (right) added SASH surfactant

Figure 4 shows fibre diameter as a function of concentration of PAN in solution with and without SASH surfactant, for fibres before and after carbonising heat treatment at 500°C. The error in the readings is taken as the standard deviation of the normal distribution of fibres measured from the SEM images. The decrease in diameter of fibres with decreasing PAN concentration can be clearly seen, as can the reduction in diameter following carbonisation. The chief effect of the surfactant was to reduce or eliminate beading at lower PAN concentrations (see Figure 3). The surfactant reduces the surface tension in the solution to enable a stable fibre to be drawn more easily at low flow rates and polymer concentrations. If the surface tension in the solution is too high, the electrostatically-stretched solution breaks up into droplets [14]. The effect of SASH is to slightly increase the fibre diameter, but the elimination of the beads in the fibres is important for the synthesis of homogeneous fibres with high aspect ratio (length/diameter).

The smallest fibre diameters, which are 106 nm for polymer fibres and 49 nm for the same fibres carbonised at 500 °C, were obtained for the 3.8 wt % PAN/DMSO solution, the solution with lowest PAN concentration that could be electrospun stably.

PAni did not fully dissolve in DMSO, but a mixture of PAN/PANI/DMSO solution did electrospin, though it was difficult to achieve stability of the spinning jet. Using a small quantity of PAni (0.6wt%) with larger quantity of PAN (3.6 wt% ) in DMSO to form a total of 4.2 wt% PAN/PAni solution had the effect of eliminating the beads in the fibres, without SASH. The diameters obtained for this solution were found to be 114 nm and 57 nm for polymer and carbon fibres respectively. The conductivity of the 4.2 wt% polymer solution increased from 12.08 μS/cm in pure PAN/DMSO solution to 15.33 μS/cm in PAni/PAN/DMSO. The addition of NaCl to the 4.2 wt% PAN/DMSO solution had the effect of reducing the beading in fibres. This is expected, because the increase in the conductivity of the solution enables the electrostatic force to overcome the surface tension. It is therefore possible that the effect of the SASH in reducing fibre beading comes from its effect of increasing the conductivity of the solution in addition to the decreased surface tension [14]. A comparison of fibre diameters of 4.2% PAN in DMSO solution, spun with different additives and carbonised, is shown in Table 1.

| Spinning solution and additives | Polymer fibres | | Carbonised at 500°C | |
|---|---|---|---|---|
| | Fibre diameter [nm] | Standard deviation [nm] | Fibre diameter [nm] | Standard deviation [nm] |
| 4.2% PAN in DMSO | 130 | 50 | 60 | 20 |
| plus 5mg/ml SASH | 170 | 80 | 110 | 40 |
| plus 10mg/ml SASH | 290 | 60 | 200 | 60 |
| plus 1mg/ml NaCl | 160 | 50 | 90 | 40 |
| plus 1mg/ml NaCl and 5 mg/ml SASH | 200 | 70 | 140 | 60 |

**Table 1:** Average fibre diameters for 4.2% PAN in DMSO with NaCl and SASH additives, as-spun polymer and after carbonisation at 500°C under vacuum

### 3.2 Graphitisation

SEM images of fibres from 6.8% PAN in DMSO solution after heat treatment (Figure 5 (a)-(d)), show how the surface morphology of the fibres change with increasing temperature. While the fibres at relatively lower temperatures (<1500 °C, Figure 5 (a) and (c)) have a smooth surface, those at higher temperatures (e.g. 2200°C, Figure 5 (b) and (d)) develop a more rough and ridged morphology with increasing temperature. This could be explained by the transformation of an amorphous carbon structure to a more crystalline graphite structure. The non-oxidised fibres, as well as having a smaller average fibre diameter, have a more uneven morphology along the fibre axis than the oxidised fibres. This result is not surprising considering the effect of the stabilisation step on the development of the fibre structure. While those fibres stabilised under an oxidising atmosphere develop a more rigid ladder structure preventing mass loss, those that are not oxidised are prone to loss of carbon through the evaporation of volatiles, as described earlier.

## 3.3 Structural Characterisation

The fibre diameters and sample mass are found to decrease with increasing heat treatment temperature, as shown in Figure 6. Oxidised fibres retain a greater proportion of their diameter and overall mass loss to 1500°C is approximately 70% of the starting polymer mass, which remains constant to 3000°C. Non-oxidised fibres lose upwards of 90 % mass to 3000°C.

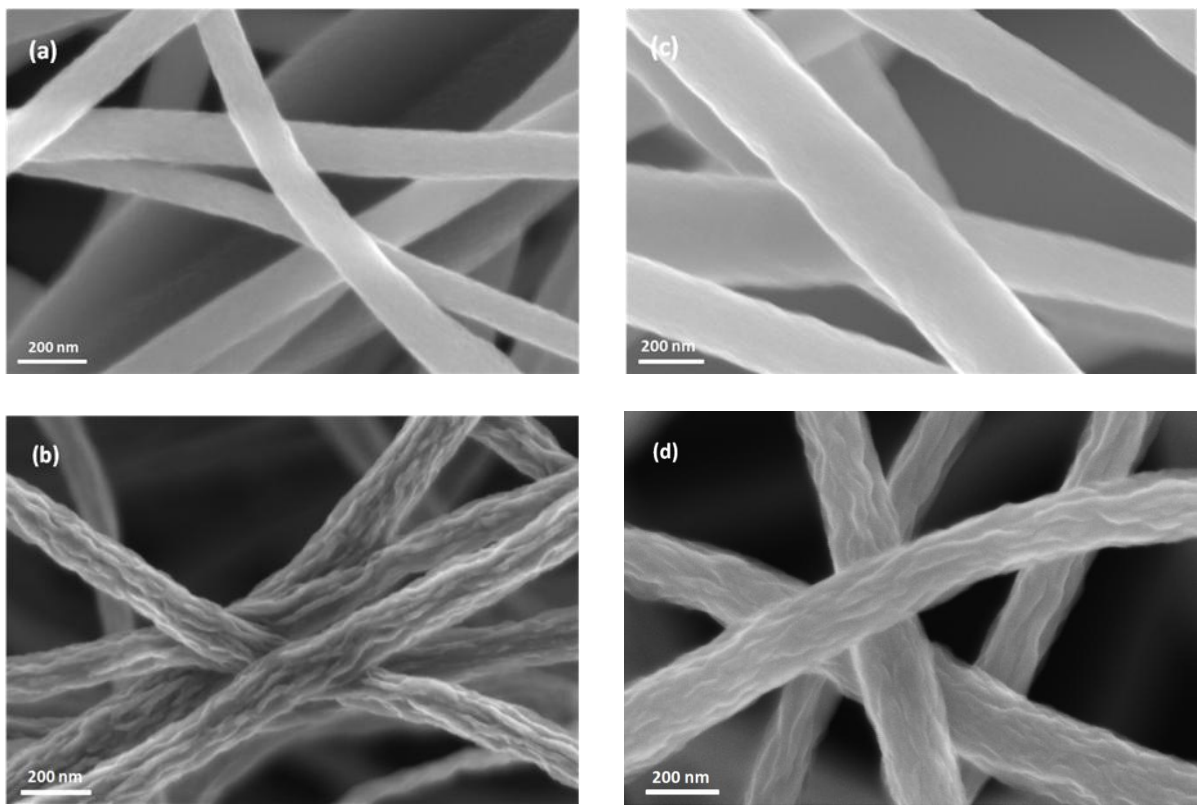

**Figure 5:** SEM images of heat treated non-oxidised fibres: a) HTT=1500°C, b) HTT=2800°C, and oxidised fibres c) HTT= 1500°C and d) HTT= 2800°C

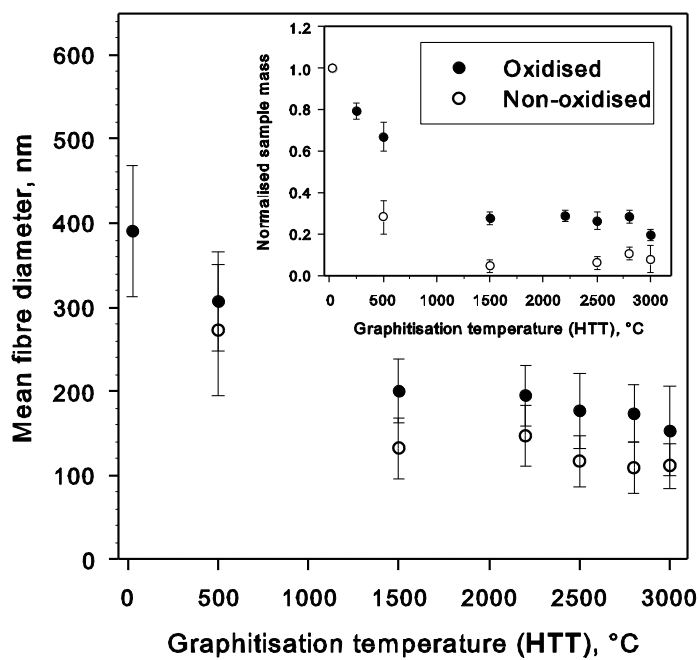

**Figure 6:** Fibre diameters and normalised mass (inset) of fibres from 6.8% PAN in DMSO solution as a function of HTT. Closed circles = oxidised fibres, open circles = non-oxidised fibres. Fibres from 6.8 wt % PAN in DMSO solution. Mass is normalised to the pre-heat treatment polymer mass for each sample.

Transmission electron microscope (TEM) images of fibres heat treated at 2800°C are shown in Figure 7. Figure 7 (a) and (b) are images of a non-oxidised fibre; (c) is that of an oxidised fibre at the same 400k magnification as (b). Though it is difficult to make a general comparison between the structure of oxidised and non-oxidised fibres with observation of only part of a fibre, the graphitic structure appears to be similar in both types of fibres. This is confirmed by XRD measurements discussed below. Both oxidised and non-oxidised fibres have an entangled ribbon structure formed of micro-fibril layers of graphite basal planes ($sp^2$ type carbon). The average layer spacing measured from the images is 3.34 Å. These ribbons appear to pass smoothly from one domain of stacking to the other, with a general orientation along the fibre axis. This observation is in agreement with the 'ribbon structure model' suggested by Diefendorf and Tokarsky for formation of graphitic structure in carbon fibres [14].

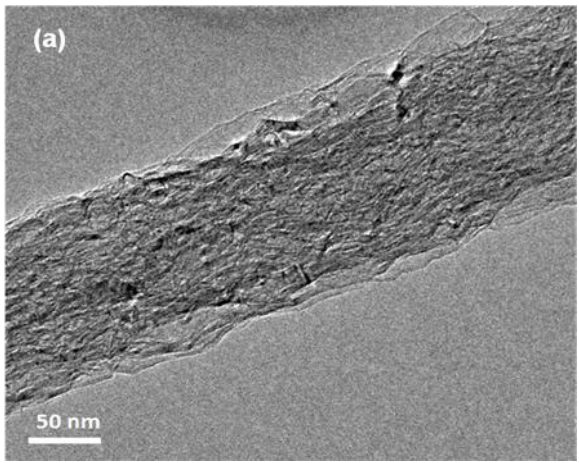

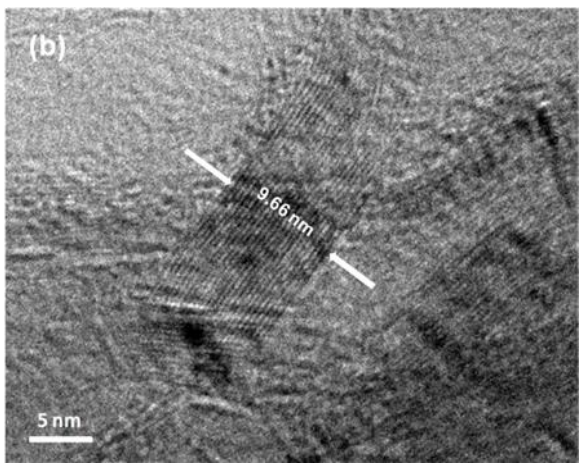

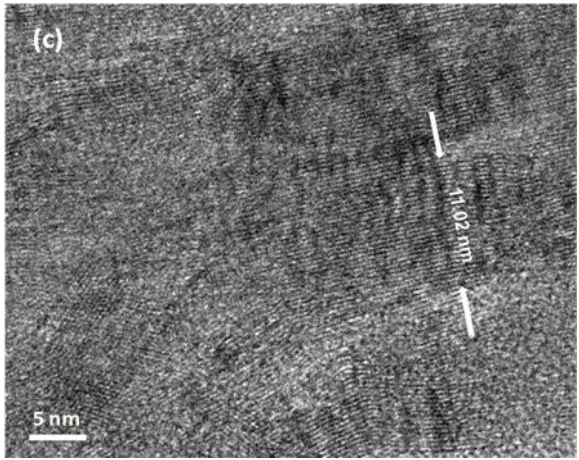

**Figure 7:** TEM images of fibres heat treated at 2800°C; (a) and (b) non-oxidised fibres; (c) oxidised fibres ((b) and (c) magnification 400k)

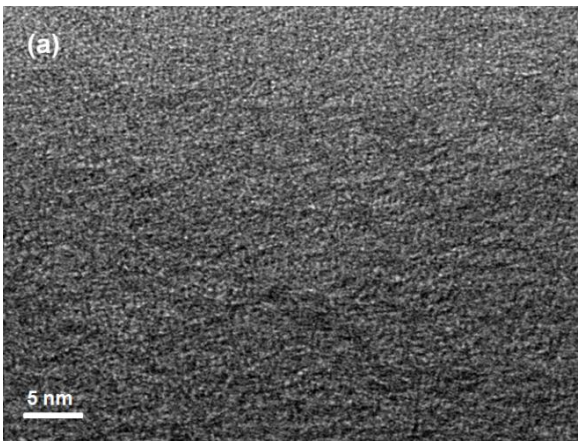

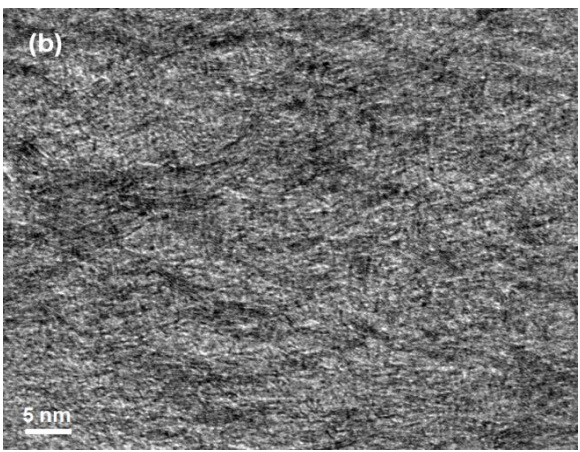

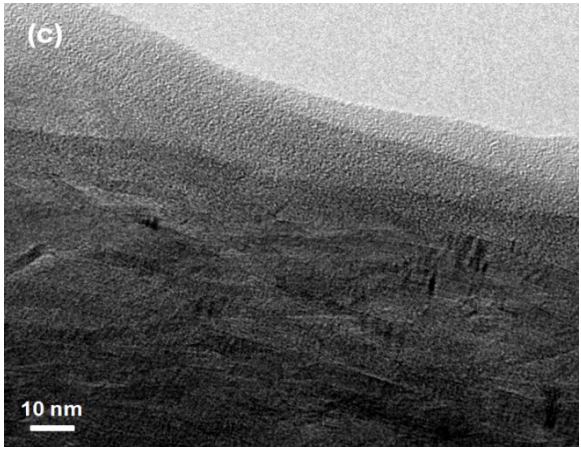

**Figure 8:** TEM images of fibres at a) HTT = 1500, b) HTT= 2200, and c) HTT = 2800°C

Figure 8 shows the development of the graphite fibrils with HTT. Both the fibril depths, which correspond to layers of graphite basal planes in the *c*-direction, and lengths increase with temperature. The orientation of (002) planes along the fibre axis also increases with heat treatment temperature, in agreement with previous work [24, 25]. It is however not clear how the fibrils interact in 3D to give the undulations and entanglement seen in the TEM images. Bennett *et al.* [26] suggest a structure with graphite layer planes interlinked both longitudinally and laterally in a complex manner. Although such a structure results in some 2D ordering, 3D ordering is not attained.

**3.4 XRD Analysis**

The background-subtracted X-ray diffraction data are presented in Figure 9. The principal graphite diffraction peaks seen for high HTT are the (002), (100), (004), (110) and (006). There is no unambiguous evidence for the presence of (*hkl*) peaks, and so the graphitic nanofibres do not appear to contain 3D ordering. Thus the local structure of the graphitic fibrils appears to be unregistered, or turbostratically disordered, stacked graphene planes.

The data show a strong signature of developing graphitisation in the *c*-axis direction. For both oxidised and non-oxidised fibres the degree of graphitisation increases with temperature, as is evident from the growing intensity of the (002) peak at 3.3 Å. A smaller growth is observed in the (004) peak at 1.7 Å. The asymmetry of the (100) peak at 2.1 Å, with a tail to the low-d side, is typical of turbostratic graphite [27]; but the (002) peak is asymmetric with a tail in the opposite direction. This is unexpected for a single phase peak; however, theory shows that when there are very few graphene layers the (002) peak is shifted towards lower d-spacings even though the interlayer spacing is unaltered. So we believe that the tail is actually due to the presence of graphite with only a few layers. Following Fujimoto *et al.* [27], it is possible to fit to a model containing a percentage of graphene bilayers, tri-layers etc. However, it was not possible to get a stable fit using this method since above 3 or 4 layers the peaks are too similar in width and position. We chose a simplified model where we fitted the (002) peak to two Lorentzians: a broad one at low d-spacings to represent the fraction of the graphite with only 2 or 3 layers and one at higher d-spacings to represent 4-layers and above. Although not strictly accurate this model produces a stable fit and gives a good indication of the development of the graphite as a function of heat treatment temperature.

The areas of the peaks are expressed as a proportion of the overall fitted peak area at each temperature in Figure 10. As the HTT increases, the proportion of two and three-layer graphite falls and many-layer graphite increases, above 80% for oxidised fibres and 60% for non-oxidised fibres. The fitted peak widths for the graphitic peak were used to determine approximate average crystallite dimensions using the Scherrer equation with a pre-factor of 0.9 [27] in the *c*-direction for both oxidised and non-oxidised GNFs, and are listed in Table 2.

The fitted graphite peak width reduces slightly between 2200 and 3000°C, although the intensity grows for both oxidised and non-oxidised fibres, suggesting that the many-layer *c*-axis crystallite nucleates at or below 2200°C and further increase in layer stacking scales only weakly with temperature. At a HTT of 2800°C and above, the c-axis crystallite dimension is ~30 graphite layers for oxidised GNFs and ~40 graphite layers for non-oxidised GNFs. This compares with the TEM images (e.g. Figure 7(c)) where up to 30 graphite layers can be counted in visible crystallite fibrils.

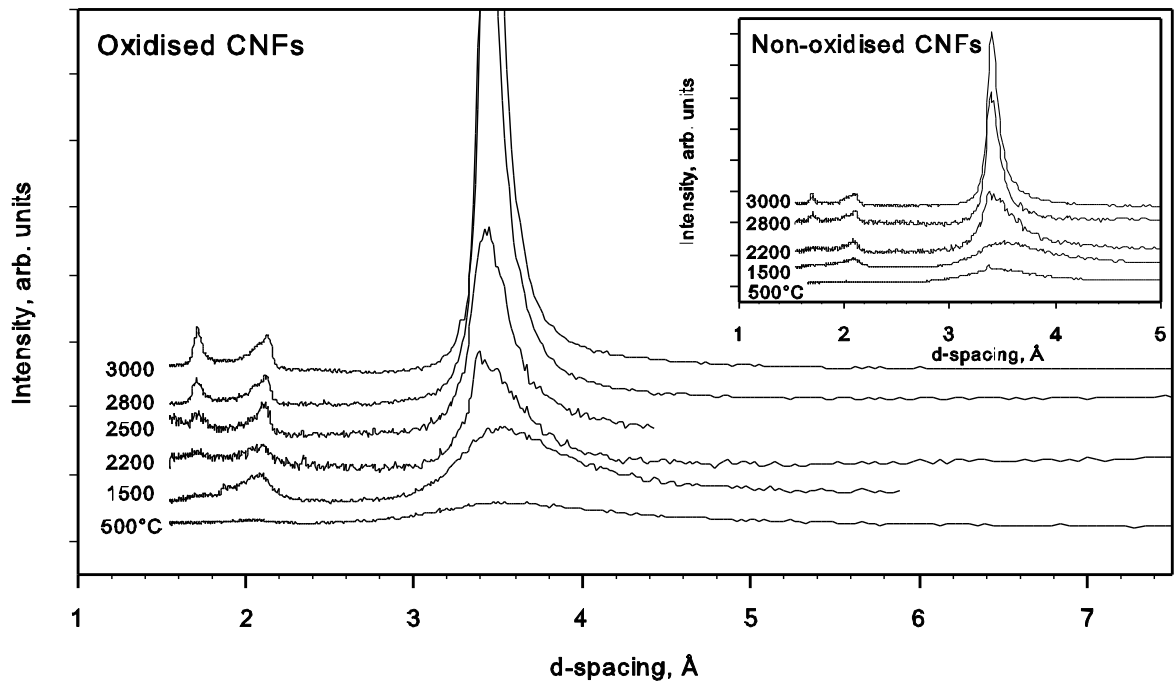

**Figure 9:** X-ray diffraction data from oxidised and (inset) non-oxidised fibres at HTT from 500 to 3000°C using Cu-Kα source. The background from the glass slide has been subtracted from each pattern.

|  | Oxidised fibres | Non-oxidised fibres |
| --- | --- | --- |
| Temp (°C) | No. of layers | No. of layers |
| 2200 | 27±3 | 29±4 |
| 2500 | 19±2 | -- |
| 2800 | 29±1 | 39±4 |
| 3000 | 28±3 | 36±1 |

**Table 2:** Peak width and resulting *c*-axis crystallite dimensions for graphitic component of (002) peak fit.

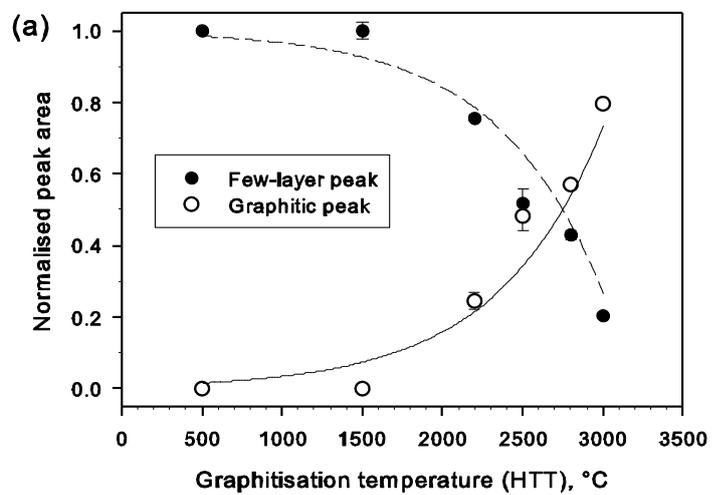

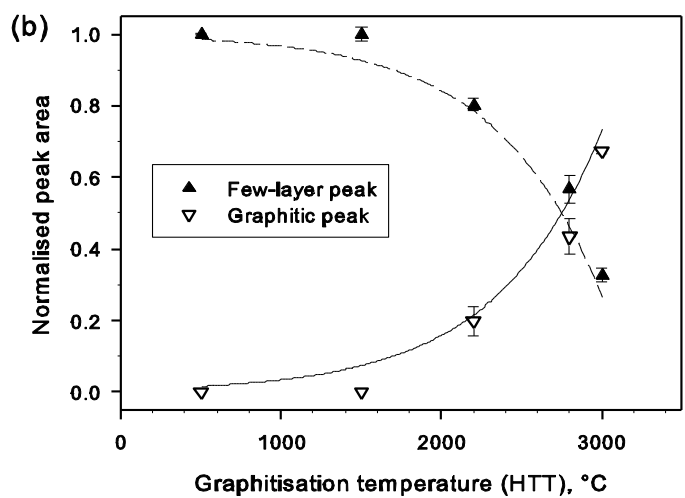

**Figure 10:** Proportional sub-peak fits to the (002) peak showing the growth of the many-layer graphitic phase from a phase consisting of few-layer (2- and 3-layer stacked) graphite with increase in HTT; (a) for oxidised and (b) non-oxidised GNF. Exponential fits are provided as a guide to the eye.

For assessing the in-plane graphite lattice dimension, the (100) peak width was measured, and the pre-factor to Scherrer's equation obtained from [27]. For oxidised GNFs with a HTT of 3000°C, the in-plane lattice dimension was found to be 5.8±0.2 nm, using a pre-factor of 1.70. For non-oxidised GNFs with the same HTT, the equivalent dimension was 4.5±0.1 nm, using a pre-factor of 1.55. This is almost certainly not a reflection of the ribbon width, but an indication of the length scale over which they are flat [28].

## 4. Conclusion

Solutions of PAN in DMSO represent a novel means of synthesising homogenous polymer nanofibres which are suitable for graphitisation through heat treatment, with or without an intermediate stabilisation step in which oxidation occurs. The electrospinning process is sensitive to the viscosity and conductivity of the solutions, and addition of surfactant reduces the beading found at low polymer concentrations, allowing the lowest polymer fibre diameters obtained. 3.8% by weight is the lower limit of PAN concentration found to produce fibres. We interpret the data as a sample containing discrete contributions from phases of varying degrees of graphitisation from bilayers to ~40 layers for non-oxidised GNFs at 3000°C, and fit the asymmetric XRD (002) peak to two subpeaks, representing few-layer and many-layer graphite. Peak fitting shows the graphitisation degree for oxidised GNFs to be slightly lower, although more of the sample mass is retained by oxidisation.

## Acknowledgments

The authors wish to thank Chris Howard and Gadipelli Srinivas (UCL) for assistance with XRD. This work was partially supported by the EU under the NEST FERROCARBON project (CEC 012881).